\newcommand{\zlabel}[1]{\label{#1} }
\newcommand{\mathbb}{\mathbf} 
\newcommand{\fc}{\frac} 
\newcommand{\lt}{\left} 
\newcommand{\rt}{\right} 
\newcommand{\mr}{\mathbf{r}}

\newcommand{\bv}{\mathbf{v}}
\newcommand{\bu}{\mathbf{u}}
\newcommand{\bw}{\mathbf{w}}

\documentclass[pra,preprint,preprintnumbers,amsmath,amssymb]{revtex4}

\usepackage{graphicx}
\usepackage{dcolumn}
\usepackage{epic}%
\usepackage{eepic}%
\usepackage{color}
\usepackage{bm}
\usepackage{comment}

\begin{document}

\preprint{APS/123-QED}

\title{Refined Madelung Equations}

\author{James P. Finley}
\email{james.finley@enmu.edu}
\affiliation{
Department of Physical Sciences,
Eastern New~Mexico University,
Station \#33, Portales, NM 88130}
\date{\today}

\begin{abstract}
The Madelung equations are two equations that are equivalent to the one-body time-dependent
Schr\"odinger equation.
In this paper, the Madelung equation, whose gradient is an Euler equation, is refined by
introducing interpretations of functions that are shown to depend only on the real-part of the
complex-valued wavefunction.
These interpretations are extensions of functions from the recently derived generalized
Bernoulli equation, applicable to real-valued quantum-mechanical stationary states.
In particular, the velocity and pressure definitions are extended so that they depend on
the real-part of a time-dependent complex-valued wavefunction.
The Bohn quantum potential is then interpreted as the sum of two terms, one involving the
kinetic energy and the other involving the pressure.
Substituting the interpreted quantum-potential into the Madelung equation gives a refined
equation containing two kinetic energy terms, a pressure term, and the external potential.
It is easily demonstrated that the refined Madelung equation, applied to the hydrogen atom
states with a nonzero magnetic quantum number, gives a fluid velocity that contains both a
radial component and a free vortex. Hence, the fluid particles have angular momentum
and move on streamlines that terminate at infinity. 
It is also demonstrated that the two velocities from the refined Madelung equation are related:
One is the real component and the other is the imaginary component of a complex
velocity. Furthermore, an Euler equation for quantum mechanical systems is derived by taking
the gradient of the refined Madelung equation.
\end{abstract}

\maketitle

\section{Introduction}

The Madelung equations \cite{Madelung:26,Madelung:27} are two equations that are equivalent to
the one-body time-dependent Schr\"odinger equation, and these equations are very similar to the
Euler equations of fluid dynamics.
The Madelung equations provide an alternate ``perspective'' of quantum mechanics compared to the
conventional one via the Schr\"odinger equation, and the possibility of a quantum-mechanical
foundation based on the Madelung equations is investigated by Wilhelm \cite{Wilhelm} and
Sonego \cite{Sonego}.
The Madelung equations do not contain a pressure.
There are at least two extensions of the Madelung equations \cite{Sorokin,Broadbridge}.

Because of the velocity definition, the Madelung equations do not provide a reasonable model
for quantum mechanical stationary states that have real valued wavefunctions.  Such states have
a static Madelung fluid, and this is not in agreement with the usual non-zero kinetic-energy
expectation value, suggesting that a satisfactory model should have some motion.
This shortcoming is also pointed out by Wyatt \cite{B6} for quantum hydrodynamics
\cite{6,7,8,9,10,11,12,13,14,15,16,17,18,19,20,21,22}, a method that is based on the Madelung
equations.

Recently, a compressible-flow generalization of the Bernoulli equation of fluid dynamics is
shown to be equivalent to the time-independent Schr\"odinger equation for one-body stationary
states with real-valued wavefunctions \cite{Finley-Arxiv}.
The generalized Bernoulli equation describes compressible, irrotational, steady
flow with a constant specific total energy, i.e, a constant energy per mass for each fluid
elements.
For the formalism, the velocity and mass-density definitions yield a generalization of the
steady-flow continuity equation where mass is \emph{not} locally conserved, and
where the pressure is proportional to the mass creation rate per volume, yielding sources and
sinks. However, over all space, the flows conserve mass.

The generalized Bernoulli equation, mentioned above. and a generalized continuity equation
provide a fluid dynamic interpretation of a class of quantum mechanical stationary states that
is an alternative to the interpretation provided by the Madelung equations
\cite{Madelung:26,Madelung:27}.  Furthermore, the \emph{integrand} of the quantum-mechanical
expectation value of the kinetic energy is given as a sum of two terms, and one term is
interpreted as the kinetic energy per volume and the other one is the pressure.  In additional,
speed of speed of equations are derived from the generalized continuity equation and a derived
variable-mass Euler equation. Applications of the formalism is applied to a fluid (or particle)
in a one-dimensional box, the ground and first excited-state of the one-dimensional harmonic
oscillator, and the hydrogen 1s and 2s states.

In this paper, in Sec.~\ref{p5544}, the Madelung equation, whose gradient is an Euler equation,
is refined by introducing interpretations from the generalized Bernoulli equation of functions
that are shown to depend only on the real-part of the complex-valued wavefunction.
In particular, the velocity and pressure definitions, applicable to a real-valued wavefunction,
are extended so that they also hold with the real part of a time-dependent complex-valued
wavefunction.
The Bohn quantum potential \cite{Bohm:52a,Bohm:52b,B6} is then interpreted as the sum of two
terms, one involving the kinetic energy and the other involving the pressure.
Substituting the interpreted quantum potential into the Madelung equation gives a refined
equation containing two kinetic energy terms, a pressure term, and the external potential.
It is easily demonstrated that the refined Madelung equation, applied to the hydrogen atom
states with a non zero magnetic quantum number, gives a velocity that contains both a radial
component and a free vortex. Hence, the fluid particles have angular momentum and move on
streamlines the terminate at infinity.

In Appendix~\ref{p5522}, it is demonstrated that the two velocities from the refined Madelung
equation are related: One is the real component and the other is the imaginary component of a
complex velocity. Also, Appendix~\ref{p5533} derives an Euler equation for quantum mechanical
systems.

\section{Refined Madelung equations \zlabel{p5544}}

Let $\phi$ be a \emph{real-valued} eigenfunction of a one-body time-independent Schr\"odinger
equation with external potential $V$. Elsewhere \cite{Finley-Arxiv} it is shown that the
following two equations are equivalent:
\begin{gather} \zlabel{0001} 
  -\fc{\hbar^2}{2m}\nabla^2\phi + V\phi = \bar{E}\phi \\ \zlabel{0002}
  \fc12 mu^2 + p\rho^{-1} + V = \bar{E}
\end{gather}
The first equation is the one-body time-independent Schr\"odinger equation.
The second equation, with the definitions,
\begin{equation}
  \zlabel{0003a} \bu_\pm = \pm\fc{\hbar}{2m}\fc{\nabla\rho}{\rho},
\end{equation}
\begin{equation}
 \zlabel{0003b}   p = -\fc{\hbar^2}{4m}\nabla^2\rho,
\end{equation}
and $u^2 = |\bu_\pm|^2$, is a steady, compressible-flow generalization of the well known Bernoulli
equation \cite{Munson,Currie} with mass density $\rho_m = m\rho$, velocity $\bu_\pm$ and
pressure $p$. Note that there are two possible velocities $\bu_+$ and $\bu_-$, giving two
possible directions along each streamline, called uphill and downhill flow, respectively.
(Elsewhere \cite{Finley-ArxivDM}, an $N$--body generalization of (\ref{0002}), based on the $N$--body
generalization of Eq.~(\ref{0001}), is derived.)

The velocity $\bu_\pm$ and mass-density $\rho_m$ definitions above yield a generalization of the
steady-flow continuity equation, given by
\begin{equation}  \zlabel{press-2b} 
  \nabla\cdot\rho_m\bu_\pm = \mp\fc{2m}{\hbar}p
\end{equation}
where mass is \emph{not} locally conserved, and where the pressure is proportional to the mass
creation rate per volume, yielding sources and sinks.  Since mass is not conserved locally,
energy also is not conserved locally. Also, since $u^2/2$ is the specific kinetic energy, it
follows that Eq.~(\ref{0002}) (divided by the mass $m$) is a statement of the conservation of
the \emph{specific} (total) energy $\bar{E}/m$ for the fluid particles. Over all space, the
flows \emph{do} conserve mass and energy; the sources and sinks cancel.

For later use, note that the equation
\begin{equation} \zlabel{p4010}
\fc12 m u^2 + p\rho^{-1} = \rho^{-1}\lt(-\fc{\hbar^2}{2m}\phi\nabla^2\phi\rt) 
\end{equation}
follows from comparing (\ref{0001}) and (\ref{0002}). However, it is easily demonstrated that
this equation, with $R$ replacing $\phi$, is an equality holding for any real valued function
$R$ such that $R^2 = \rho$. In the derivation below, we extend the definitions (\ref{0003a})
and (\ref{0003b}) to the cases where $R$ is the real part of a time-dependent complex-valued
wavefunction $\psi$.

The one-body time-dependent Sch\"odinger equation is \cite{Raimes,Levine}
\begin{equation} \zlabel{p1010}
  i\hbar  \partial \psi = -\fc{\hbar^2}{2m}\nabla^2 \psi + V\psi
\end{equation}
where $[\partial\psi](t) = \partial\psi/\partial t$.
The two Madelung equations, given by \cite{Madelung:26,Madelung:27,B6} 
\begin{gather}  \zlabel{p4444}
  \partial \rho_m + \nabla\cdot(\rho_m\bv) = \mathbf{0}
  \\ \zlabel{p4445}
  -\partial S  = \fc{1}{2m}\nabla S\cdot\nabla S + Q + V 
\end{gather}
are equivalent to the Sch\"odinger equation, where the wavefunction ansatz is
\begin{equation} \zlabel{p2222}
\psi(x,t) = R(x,t)e^{iS(x,t)/\hbar},
\end{equation}
and $R$ is required to be nonnegative. The function $Q$, defined below, is the Bohmian
quantum potential. The Madelung equations characterize the flow of a
compressible gas with mass density $\rho_m = m|\psi|^2$ and velocity
\begin{equation} \zlabel{p4182}
\bv = \fc{\nabla S}{m}
\end{equation}
Eq.~(\ref{p4444}) is the continuity equation, a statement of the conservation of mass. The
gradient of (\ref{p4445}) is very similar to the Euler equation of fluid dynamics.

Using the above velocity definition with $v^2 = \bu_\pm\cdot\bu_\pm$, (\ref{p4445}) can be written
\begin{equation} \zlabel{p2050}
   -\partial S  = \fc{1}{2}m v^2  + Q + V
\end{equation}

The Bohm quantum potential $Q$ is defined by \cite{Bohm:52a,Bohm:52b,B6} 
\[
Q = -\fc{\hbar^2}{2m}\fc{1}{R}\nabla^2 R
\]
Using $R = \rho^{1/2}$ we have
\begin{gather*}
  \nabla^2R = \nabla\cdot\nabla\rho^{1/2} = \fc12\nabla\cdot(\rho^{-1/2}\nabla\rho)
  =  \fc12\nabla\rho^{-1/2}\cdot \nabla\rho  + \fc12\rho^{-1/2}\nabla^2\rho
  \\ =  -\fc14\rho^{-3/2}\nabla\rho\cdot \nabla\rho  + \fc12\rho^{-1/2}\nabla^2\rho \hspace{15ex}
\end{gather*}
Hence,
\[
Q = \fc{\hbar^2}{8m}\rho^{-2}\nabla\rho\cdot \nabla\rho  - \fc{\hbar^2}{4m}\rho^{-1}\nabla^2\rho \hspace{15ex}
\]
Next, as discussed above, we extend the velocity and pressure definitions, (\ref{0003a}) and
(\ref{0003b}), so that they hold for $\phi = R$, where $R$ is the real part of the complex
valued wavefunction $\psi$, and note that we still have $R^2 = \rho$. Doing this, and using
Eq.~(\ref{p4010}), we obtain
\begin{equation}
Q = \fc12 m u^2 + p\rho^{-1} = \rho^{-1}\lt(-\fc{\hbar^2}{2m}R\nabla^2R\rt), 
\end{equation}
where $u^2= \bu\cdot\bu$.  Substituting for $Q$, (\ref{p2050}) becomes the desired result:
\begin{equation} \zlabel{p2055}
   -\partial S  = \fc{1}{2}m v^2  + \fc12 m u^2 + p\rho^{-1} + V 
\end{equation}
This equation is a refinement of (\ref{p2050}), containing two kinetic energy terms, a
``compression'' energy term $p\rho^{-1}$, and the external potential $V$.  The right-hand-side
of this equation (divided by $m$) can be interpreted as the time dependent total specific
energy, i.e., a Hamiltonian function for specific energy.

If $\psi$ is a stationary state then $\psi(x,t) = R(x)e^{-iEt/\hbar}$, so $S(t) = -\bar{E}t$, giving
\begin{equation} \zlabel{p2058}
\fc{1}{2}m v^2  + \fc12 m u^2 + p\rho^{-1} + V  = \bar{E}
\end{equation}
This equation is a generalization of (\ref{0002}) holding also for complex-valued one-body stationary states.

For real valued stationary states, since the velocity $\bu$ satisfies (\ref{0003a}), the
directions of $\bu_\pm$ is perpendicular to the level surfaces of $\phi$ and $\rho$.  Hence,
for the hydrogen-atom real-valued states, the streamlines terminate at points at infinity. This
behavior also holds for the streamlines from the velocity field $\bu$ alone, for steady-state
complex-valued hydrogen wavefunctions, ignoring $\bv$.  In spherical coordinate, the hydrogen
wavefunctions can be written $\psi(r,\theta,\phi) = R(r,\theta)e^{\pm i\ell\phi}$
\cite{Raimes,Levine,Bransden}, where $\ell$, an integer, is the magnetic quantum number.  From
the ansatz (\ref{p2222}), we have $S = \pm\hbar\ell\phi$.  Hence $\bv = \pm h\ell
(r\sin\theta)^{-1} \mathbf{\hat{\phi}}$, so, for $\ell\ne 0$, the flows have a vortex, and the
vortex is a free vortex, since $\nabla\times\bv = \mathbf{0}$.  Therefore, the hydrogen atom
states with $\ell \ne 0$ have fluid particle with a nonzero angular momentum, and this has some
agreement with the angular momentum predictions from quantum mechanics, but the model has
nothing to say about the measurement process. Also, using the gradient in spherical
coordinates, it follows that $\bu\cdot\bv= \mathbf{0}$, giving $|\bu + \bv|^2 = u^2 + v^2$.

\appendix

\section{Equalities for the Velocities \zlabel{p5522}}

Next we show that
\begin{equation} \zlabel{p2880}
  \bu_\pm = \pm\text{Re}\lt(\fc{\hbar}{m}\fc{\nabla\psi}{\psi}\rt),
  \quad \bv = \text{Im}\lt(\fc{\hbar}{m}\fc{\nabla\psi}{\psi}\rt)
\end{equation}
The first one is obtained by substituting $\rho = R^2$ into (\ref{0003a}):
\[
\bu_\pm = \pm\fc{\hbar}{2m}\fc{\nabla R^2}{R^2} = \pm\fc{\hbar}{m}\fc{\nabla R}{R}
= \pm\text{Re}\lt(\fc{\hbar}{m}\fc{\nabla\psi}{\psi}\rt)
\]
The second one is proven in the following sequence, starting with
the ansatz $\psi= Re^{iS/\hbar}$ and requiring (\ref{p4182}) for $\bv$ to hold.
\begin{gather*}
  \nabla\psi = (\nabla R) e^{iS/\hbar} + i\hbar^{-1}Re^{iS/\hbar}\nabla s  \\
  \fc{\nabla\psi}{\psi} = (\nabla R)R^{-1} + i\hbar^{-1} \nabla s  \\
  \fc{\hbar}{m}\fc{\nabla\psi}{\psi} = \fc{\hbar}{m}(\nabla R)R^{-1} + i\fc{\nabla s}{m}  \\
  \mathbf{v} = \fc{\nabla s}{m} = \text{Im}\lt(\fc{\hbar}{m}\fc{\nabla\psi}{\psi}\rt) 
\end{gather*}
Let $\bw = \hbar \nabla\psi/(m\psi)$. Equations~(\ref{p2880}) implies that $\bw = \bu + i\bv$,
and $|\bw|^2 = u^2 + v^2$.

\section{The Euler equation for quantum mechanical systems \zlabel{p5533}}

For a fluid of classical mechanics, the Euler equation with variable mass and steady flow is 
\cite{Finley-Arxiv}
\begin{gather}
  \fc12\rho_m\nabla u^2 + \nabla\cdot(\rho_m\bu)\bu + \nabla p + \rho\nabla V = \mathbf{0}
\end{gather}
Elsewhere \cite{Finley-Arxiv} it is shown that this equation is equivalent the gradient of
(\ref{0002}), which, for $\rho(\mr) \ne 0$, can be written
\begin{gather} \zlabel{0002b}
  \fc12 \rho_m \nabla u^2 + \rho\nabla\lt(\fc{p}{\rho}\rt) + \rho\nabla V = \mathbf{0}
\end{gather}
Comparing the two, we have
\begin{equation} \zlabel{p2010}
\rho\nabla\lt(\fc{p}{\rho}\rt) = \nabla\cdot(\rho_m\bu)\bu + \nabla p
\end{equation}
Taking the gradient of (\ref{p2055}) and multiplying the result be $\rho$ we have
\begin{equation} \zlabel{p2075}
   \rho_m\partial \bv  + \fc{1}{2}\rho_m \nabla v^2  + \fc12 \rho_m \nabla u^2 + \rho\nabla\lt(\fc{p}{\rho}\rt) + \rho\nabla V = \mathbf{0}
\end{equation}
Substituting (\ref{p2010}), we obtain
\begin{equation} \zlabel{p4075}
   \rho_m\partial \bv  + \fc{1}{2}\rho_m \nabla v^2  + \fc12 \rho_m \nabla u^2 + \nabla\cdot(\rho_m\bu)\bu + \nabla p + \rho\nabla V = \mathbf{0},
\end{equation}
a Euler equation applicable to quantum mechanical systems. The other two equations being the
continuity equations, (\ref{p4444}) and (\ref{press-2b}).

\section{Summary}

In this paper, the refined Madelung equation~(\ref{p2055}) is derived. This equation with the
continuity equation~(\ref{p4444}) are equivalent to the one-body time dependent Schr\"odinger
equation~(\ref{p1010}). The two velocities, $\bu_\pm$ and $\bv$, and the pressure $p$ are
defined by Eq.~(\ref{0003a}), (\ref{p4182}) and~(\ref{0003b}).  Equation~(\ref{p2055}) can be
viewed as combination of a Madelung equation (\ref{p2050}) and the generalized Bernoulii
equation (\ref{0002}), or as a refinement of the corresponding Madelung equation with new
interpretations for a pressure $p$ and an additional velocity component~$\bu_\pm$.

\bibliography{jfinley}

\end{document}